\documentclass[10pt,a4paper,english]{article}\usepackage[]{graphicx}\usepackage[]{xcolor}
\makeatletter
\def\maxwidth{ %
  \ifdim\Gin@nat@width>\linewidth
    \linewidth
  \else
    \Gin@nat@width
  \fi
}
\makeatother

\definecolor{fgcolor}{rgb}{0.345, 0.345, 0.345}

\usepackage{framed}
\makeatletter
 {\par\unskip\endMakeFramed%
 \at@end@of@kframe}
\makeatother

\definecolor{shadecolor}{rgb}{.97, .97, .97}
\definecolor{messagecolor}{rgb}{0, 0, 0}
\definecolor{warningcolor}{rgb}{1, 0, 1}
\definecolor{errorcolor}{rgb}{1, 0, 0}
\newenvironment{knitrout}{}{} 

\usepackage{alltt}

\usepackage[hyphens]{url}
\usepackage{amsmath}
\usepackage{amsfonts}
\usepackage{amssymb}
\usepackage{amsthm}

\theoremstyle{plain}

\theoremstyle{definition}

\theoremstyle{remark}

\PassOptionsToPackage{hyphens}{url}\usepackage{hyperref}
\usepackage{hyperref}
\usepackage[square,numbers]{natbib}
\usepackage{orcidlink}
\usepackage{gitinfo2}

\usepackage[newitem,newenum,increaseonly]{paralist}


\usepackage[colorinlistoftodos]{todonotes}

\usepackage[notheorems]{SharpeR}

    \usepackage{color} 
    \usepackage{enumerate} 
    \usepackage{amsmath} 
    \usepackage{amssymb} 
		\usepackage{fancyvrb} 
    \usepackage{hyperref}

    \definecolor{orange}{cmyk}{0,0.4,0.8,0.2}
    \definecolor{darkorange}{rgb}{.71,0.21,0.01}
    \definecolor{darkgreen}{rgb}{.12,.54,.11}
    \definecolor{myteal}{rgb}{.26, .44, .56}
    \definecolor{gray}{gray}{0.45}
    \definecolor{lightgray}{gray}{.95}
    \definecolor{mediumgray}{gray}{.8}
    \definecolor{inputbackground}{rgb}{.95, .95, .85}
    \definecolor{outputbackground}{rgb}{.95, .95, .95}
    \definecolor{traceback}{rgb}{1, .95, .95}
    \definecolor{red}{rgb}{.6,0,0}
    \definecolor{green}{rgb}{0,.65,0}
    \definecolor{brown}{rgb}{0.6,0.6,0}
    \definecolor{blue}{rgb}{0,.145,.698}
    \definecolor{purple}{rgb}{.698,.145,.698}
    \definecolor{cyan}{rgb}{0,.698,.698}
    \definecolor{lightgray}{gray}{0.5}
    
    \definecolor{darkgray}{gray}{0.25}
    \definecolor{lightred}{rgb}{1.0,0.39,0.28}
    \definecolor{lightgreen}{rgb}{0.48,0.99,0.0}
    \definecolor{lightblue}{rgb}{0.53,0.81,0.92}
    \definecolor{lightpurple}{rgb}{0.87,0.63,0.87}
    \definecolor{lightcyan}{rgb}{0.5,1.0,0.83}

    \definecolor{incolor}{rgb}{0.0, 0.0, 0.5}
    \definecolor{outcolor}{rgb}{0.545, 0.0, 0.0}

    \hypersetup{
      breaklinks=true,  
      colorlinks=true,
      urlcolor=blue,
      linkcolor=darkorange,
      citecolor=darkgreen,
      }
    
\providecommand{\RMAT}[1][{}]{\MtxUL{R}{#1}{}}
\providecommand{\RHAT}[1][{}]{\MtxUL{\hat{R}}{#1}{}}

\providecommand{\makerho}[3][{\vone}]{{#2}\wrapParens{\ogram{#1}} + {#3}\,\eye}

\providecommand{\qtuk}[3]{\mathUL{q}{}{{#1},{#2},{#3}}}

\providecommand{\posthoc}{\emph{post hoc}\xspace}

\providecommand{\code}[1]{\texttt{#1}}

\IfFileExists{upquote.sty}{\usepackage{upquote}}{}
\begin{document}

\title{A \posthoc test on the \txtSR}
\author{\orcidlink{0000-0002-4197-6195} Steven E. Pav \thanks{\email{steven@gilgamath.com}.
The source code to build this document is available at
\href{http://www.github.com/shabbychef/posthoc_sr}{\normalfont\texttt{www.github.com/shabbychef/posthoc\_sr}}.
This revision was built from commit \texttt{75828acb2f3e1df37cd835310c0eb5826f9ca332} of that repo.
}}

\maketitle

\begin{abstract}
  We describe a \posthoc test for the \txtSR, analogous to Tukey's test
  for pairwise equality of means. 
  The test can be applied after rejection of the hypothesis that all
  population \txtSNRs are equal.
  The test is applicable under a simple correlation structure among
  asset returns.
  Simulations indicate the test maintains nominal type I rate under
  a wide range of conditions and is moderately powerful under
  reasonable alternatives.
\end{abstract}

\section{Introduction}

Sharpe's ``reward-to-variability ratio'' 
was originally devised to compare the performance of mutual funds.  
Sharpe found it to be weakly predictive of out-of-sample performance
when measured over a \emph{decade} of returns.  \cite{Sharpe:1966}
Early research on the \txtSR, as it came to be known\footnote{Although it was
described over a decade earlier by A. D. Roy. \cite{roySafety1952}}, 
ignored its statistical nature, treating it like an observable population parameter,
though this was soon remedied.  \cite{CambridgeJournals:4493808,jobsonkorkie1981,lo2002,pav_the_book}
More recently, statistical procedures have been proposed to test whether the
population \txtSRs of several assets (\eg mutual funds, ETFs, hedge funds, \etc)
are equal.  \cite{Leung2008,wright2014}
Here we propose a test to be used to compare pairwise differences after application
of such a test.


\section{The test}

Suppose one has observed \ssiz \iid samples of some $\nstrat$-vector \vreti,
representing the returns of $\nstrat$ different ``assets.'' We imagine these
assets to be different mutual funds, or trading strategies, ETFs, \etc
From the sample one computes the \txtSR of each asset, resulting in 
a $\nstrat$-vector, \svsr. 

One natural question to ask is whether the ``\txtSNR'' (the
population analogue of the \txtSR) of each asset is equal.
One can test the hypothesis of equal \txtSNR via the
tests of Leung and Wong, or Wright, Yam and Yung.
\cite{Leung2008,wright2014}
The test of Wright \etal, for example, uses asymptotic
normality of the \svsr to construct a statistic following a 
$\chi^2$ distribution under the null.
In the case where one rejects the null hypothesis of equality,
one seeks a \posthoc test, to determine which pairs of the \nstrat
assets have different \txtSNR.

Testing the equality of \txtSNRs is analogous to the classical
procedure for testing equality of means via ANOVA.  \cite{bain1992introduction}
The \posthoc procedure that classically followed a rejection
of the null in ANOVA is Tukey's range test,
sometimes called the ``honest significant difference'' (HSD) test.
\cite{tukey_hsd,bretz2016multiple}
In the ANOVA, and Tukey's HSD, the quantity is assumed
to have identical variance among all individuals, but
potentially different means in different groups.
For this reason, the variance is estimated by pooling all
observations.
It is unnecessary to assume equal volatility of the returns
for the $\nstrat$ different assets when testing the \txtSNR.
While this simplifies our \posthoc test somewhat, 
typically in the testing of 
asset returns one observes them contemporaneously,
and they are generally correlated.

Tukey's HSD proceeds by computing an upper quantile on the
\emph{range} of independent normals divided by a rescaled
$\chi$ variable.  When the means of two individuals
differ by more than this amount, one rejects the null
that they are equal.  
Our test will perform a similar computation.

Previously the author showed that when returns are
drawn from a multivariate normal distribution with
correlation \RMAT, then
\begin{equation}
\svsr \approx 
\normlaw{\pvsnr,\oneby{\ssiz}\wrapParens{\RMAT +
	\frac{1}{2} \Mdiag{\pvsnr} \wrapParens{\RMAT\hadm\Mtx{R}} \Mdiag{\pvsnr}}},
  \label{eqn:apx_srdist_gaussian}
\end{equation}
where \pvsnr is the vector of \txtSNRs and \ssiz is the sample size.  \cite{pav2019maxsharpe}
Note that the approximate covariance matrix here generalizes the well-known 
standard error of the scalar \txtSR.  \cite{Johnson:1940,jobsonkorkie1981,lo2002,pav_ssc}
In the case of the small \txtSNRs likely to be encountered in practice,
that approximation may be further simplified to 
\begin{equation}
\svsr \approx \normlaw{\pvsnr,\oneby{\ssiz}\RMAT}.
  \label{eqn:apx_srdist_simple}
\end{equation}
Then under the null hypothesis that $\pvsnr=\pvsnr[0]$, one observes
\begin{equation}
  \vect{z} = \sqrt{\ssiz}\ichol{\RMAT} \wrapParens{\svsr - \pvsnr[0]} \approx \normlaw{\vzero,\eye},
  \label{eqn:bonf_zform_simple}
\end{equation}
where $\ichol{\RMAT}$ is the inverse of the (symmetric) square root of $\RMAT$.

As previously, we assume a simple rank-one form for the correlation matrix,
\begin{equation}
\label{eqn:simple_RMAT}
\RMAT=\makerho{\rho}{\wrapParens{1-\rho}},
\end{equation}
where $\abs{\rho} \le 1$. \cite{pav2019maxsharpe}
Under this assumption, it is simple to show that
\begin{equation}
  \label{eqn:simple_RMAT_ichol_simple}
  \ichol{\RMAT} = 
   \makerho{c}{\wrapParens{1-\rho}^{-1/2}},
\end{equation}
for some constant $c$.

Now we consider the difference in \txtSRs of two assets, indexed by $i$ and $j$.
Let $\vect{v}=\basev[i]-\basev[j]$, where $\basev[i]$ is the \kth{i} column of
the identity matrix. From \eqnref{bonf_zform_simple} we have
\begin{align*}
  \trAB{\vect{v}}{\vect{z}} &= \trAB{\vect{v}}{\sqrt{\ssiz}\ichol{\RMAT} \wrapParens{\svsr - \pvsnr[0]}},\nonumber\\
  &= \sqrt{\ssiz}\trAB{\vect{v}}{\wrapBracks{\makerho{c}{\wrapParens{1-\rho}^{-1/2}}}} \wrapParens{\svsr - \pvsnr[0]},\\
  &= \sqrt{\frac{\ssiz}{1-\rho}}\trAB{\vect{v}}{\svsr}.
\end{align*}
Here we have used that $\trAB{\vect{v}}{\vone} = 0$ and 
under the null hypothesis, \pvsnr[0] is some constant times \vone. Thus
\begin{equation}
  \ssr[i] - \ssr[j] = \sqrt{\frac{1-\rho}{\ssiz}} \wrapParens{z_i - z_j}.
\end{equation}
Now note that the \vect{z} is distributed as a standard multivariate normal. 
So the \emph{range} of \svsr, which is to say $\max_{i,j}\wrapParens{\ssr[i] - \ssr[j]}$,
is distributed as $\sqrt{\wrapParens{1-\rho}/\ssiz}$ times the range of
a standard $\nstrat$-variate normal.

To quote this as a hypothesis test, 
\begin{equation}
  \label{eqn:hyp_test_inf_df}
  \max_{i,j} \abs{\ssr[i] - \ssr[j]} \ge HSD = \qtuk{1-\typeI}{\nstrat}{\infty} \sqrt{\frac{(1-\rho)}{\ssiz}},
\end{equation}
with probability \typeI, 
where the $\qtuk{1-\typeI}{k}{l}$ is the upper $\typeI$-quantile 
of the Tukey distribution with $k$ and $l$ degrees of freedom. 
In the \textsc{R} language, this quantile may be computed via the
\texttt{qtukey} function.  \cite{Rlang,OdehEvans}
With $l=\infty$, the cutoff $HSD$ is the rescaled upper \typeI
quantile of the range of \nstrat independent Gaussians.
That is, $\qtuk{1-\typeI}{\nstrat}{\infty}$ is the number such that
$$
1-\typeI = \nstrat \int_{-\infty}^{\infty} \dnorm[x]\wrapParens{\pnorm[x + \qtuk{1-\typeI}{\nstrat}{\infty}] - \pnorm[x]}^{\nstrat-1} \dx.
$$

We note that the approximation of \eqnref{apx_srdist_gaussian} may be too coarse
for the computation of the $HSD$ cutoff. 
Even if the covariance given there is approximately correct, it is likely
that the distributional shape of $\svsr$ is far enough from multivariate normal
that we cannot use Tukey's distribution for a cutoff, especially when $\ssiz$ is
small and $\nstrat$ is large.
In that case, one is tempted to heuristically compare the observed range to
\begin{equation}
\label{eqn:hsd_n_df}
  HSD=\qtuk{1-\typeI}{\nstrat}{\ssiz-1}\sqrt{\frac{(1-\rho)}{\ssiz-1}}.
\end{equation}
The reasoning here is that we are essentially computing the range of (non-independent)
$t$ statistics, up to scaling, which is almost the same as the Tukey distribution,
which is the ratio of the range of normals divided by a pooled $\chi$ variable.
In our testing below we will refer to the cutoff of \eqnref{hyp_test_inf_df} as 
``$df=\infty$'' and the cutoff of \eqnref{hsd_n_df} as the ``$df=\ssiz-1$'' cutoff.

\paragraph{Bonferroni Cutoff: }

We note that an alternative calculation provides a very similar cutoff value.
Considering two assets with correlation $\rho$.
Suppose the \txtSNRs of the two assets are, respectively, 
$\psnr\wrapParens{1 + \epsilon}$ and \psnr.
The difference in \txtSRs can then be shown to be approximately normal: \cite{pav_ssc}
\begin{equation}
\wrapBracks{\ssr[1] - \ssr[2]}
\rightsquigarrow 
\normlaw{\epsilon\psnr,\frac{2}{\ssiz} \wrapParens{1 - \rho} +
\frac{\psnrsq}{2\ssiz}\wrapParens{1 + \wrapParens{1+\epsilon}^2 - 2\rho^2\wrapParens{1+\epsilon}}
}.
\label{eqn:del_correlated_sr}
\end{equation}
Assuming that $\psnrsq / \ssiz$ will be very small for most practical work, one can
compute the alternative cutoff, a ``Bonferroni Cutoff,'' as 
$$
BC = \sqrt{\frac{2 \wrapParens{1-\rho}}{\ssiz}} \qnorm{1 - \typeI/{\nstrat \choose 2}},
$$
where $\qnorm{\typeI}$ is the $\typeI$ quantile of the standard normal distribution.
This cutoff is based on a Bonferroni correction that recognizes we are performing
$\nstrat \choose 2$ pairwise comparison tests. 
The cutoff $BC$ is typically very
similar to $HSD$ (for $df=\infty$) or slightly smaller (and is easier to compute).
We note that since $BC$ is based on a normal approximation, it may suffer
from the same issues that the $HSD$ cutoff does for small samples. 
However, there is hope one can compute an exact small $\ssiz$ Bonferroni cutoff.

\paragraph{Arbitrary correlation structure: }

The test outlined above is strictly only applicable
to the rank-one correlation matrix, 
$\RMAT=\makerho{\rho}{\wrapParens{1-\rho}}$.
To apply the test to assets with arbitrary correlation matrices,
one would like to appeal to a stochastic dominance result.
For example, if one could adapt Slepian's lemma to the distribution
of the \emph{range}, then the above analysis could be applied where $\rho$
is the smallest off-diagonal correlation, to give a test with maximum
type I rate of $\typeI$.
However, it is not immediately clear that Slepian's lemma can be so
modified.
\cite{slepian1962one,zeitouni2015gaussian,yin2019stochastic}
The Bonferroni Cutoff, however, is easily adapted to this kind of
worst-case analysis, however.


\section{Examples}

\subsection{Simulations under the null}

\paragraph{Basic Simulations}

We spawn 4 years of daily data (252 days per year) from 
16 assets, each with \txtSNR of $1\yrto{-\halff}$.
Returns are multivariate normal with correlation
$\RMAT=\makerho{\rho}{\wrapParens{1-\rho}},$ for
$\rho=0.8$. 
We compute the \txtSR of the simulated returns, $\svsr$, then compute
the range $\max_{i,j} \ssr[i] - \ssr[j]$.
We repeat this experiment 5,000 times.
In \figref{null_basic_qq_plot}, we Q-Q plot these simulated ranges of 
the \txtSR against the theoretical quantile function
$$
\sqrt{\frac{1-\rho}{\ssiz}}\qtuk{\cdot}{\nstrat}{\infty}.
$$
We see good agreement between theoretical and actual, with
little deviance from the $y=x$ line.

\begin{knitrout}\small
\definecolor{shadecolor}{rgb}{0.969, 0.969, 0.969}\color{fgcolor}\begin{figure}[h]
\includegraphics[width=0.975\textwidth,height=0.646\textwidth]{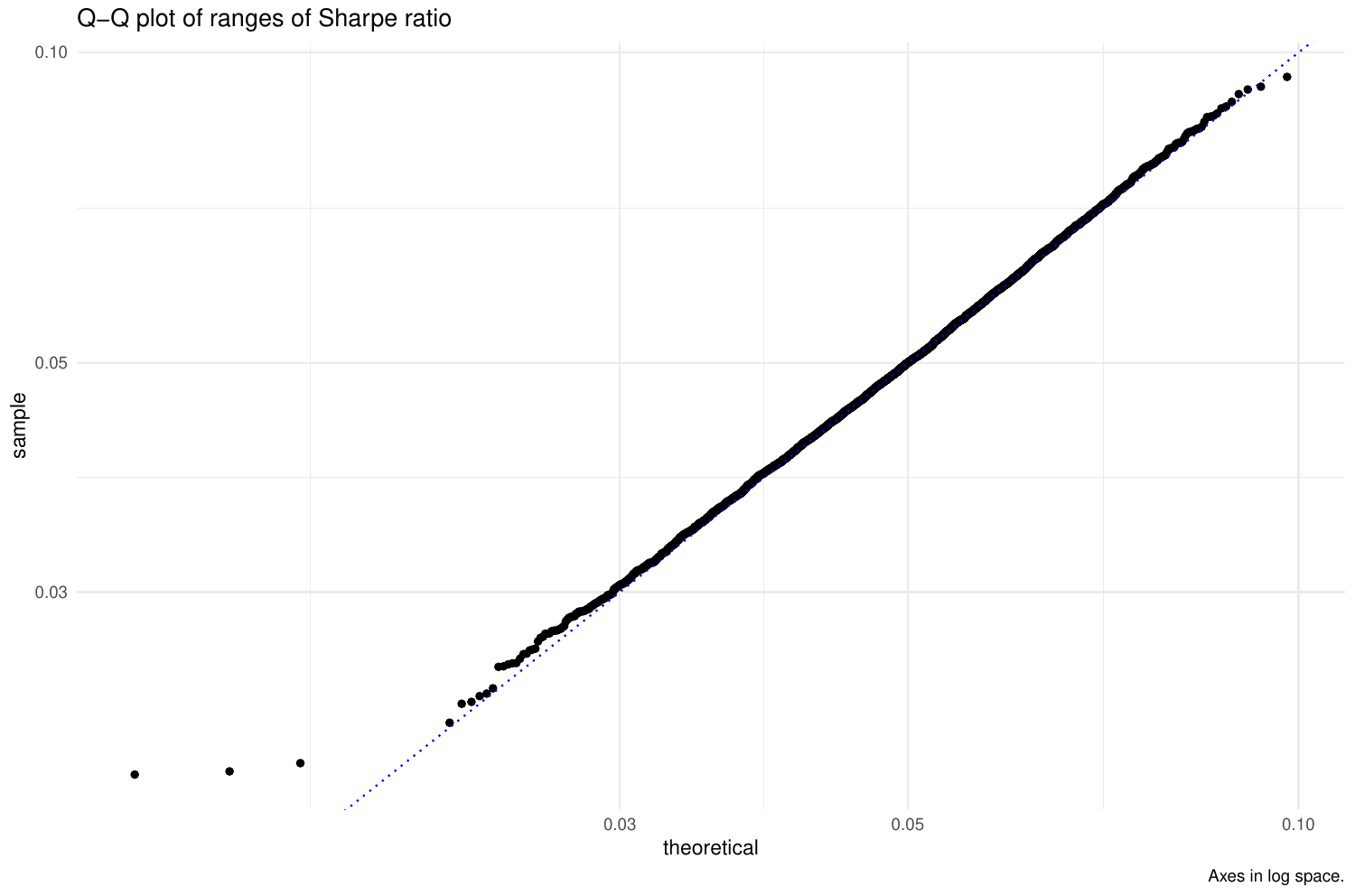} \caption[The quantiles of the range of \txtSR from 5,000 simulations are plotted against a transformed Tukey distribution, $\sqrt{(1-\rho)/\ssiz}\, \qtuk{\cdot}{\nstrat}{\infty}$]{The quantiles of the range of \txtSR from 5,000 simulations are plotted against a transformed Tukey distribution, $\sqrt{(1-\rho)/\ssiz}\, \qtuk{\cdot}{\nstrat}{\infty}$. The points show little deviation from the plotted $y=x$ line. }\label{fig:null_basic_qq_plot}
\end{figure}

\end{knitrout}

We then convert these simulated ranges to p-values via the
\texttt{ptukey} function in \textsc{R}, using the
$df=\infty$ cutoff and the actual $\rho$.
We Q-Q plot these putative p-values against a uniform law in
\figref{null_basic_pp_plot}, and again find very good agreement.
To check the tails we have transformed the p-values to
$p \mapsto \abs{2p - 1}$ and plotted in log-log scale.

\begin{knitrout}\small
\definecolor{shadecolor}{rgb}{0.969, 0.969, 0.969}\color{fgcolor}\begin{figure}[h]
\includegraphics[width=0.975\textwidth,height=0.646\textwidth]{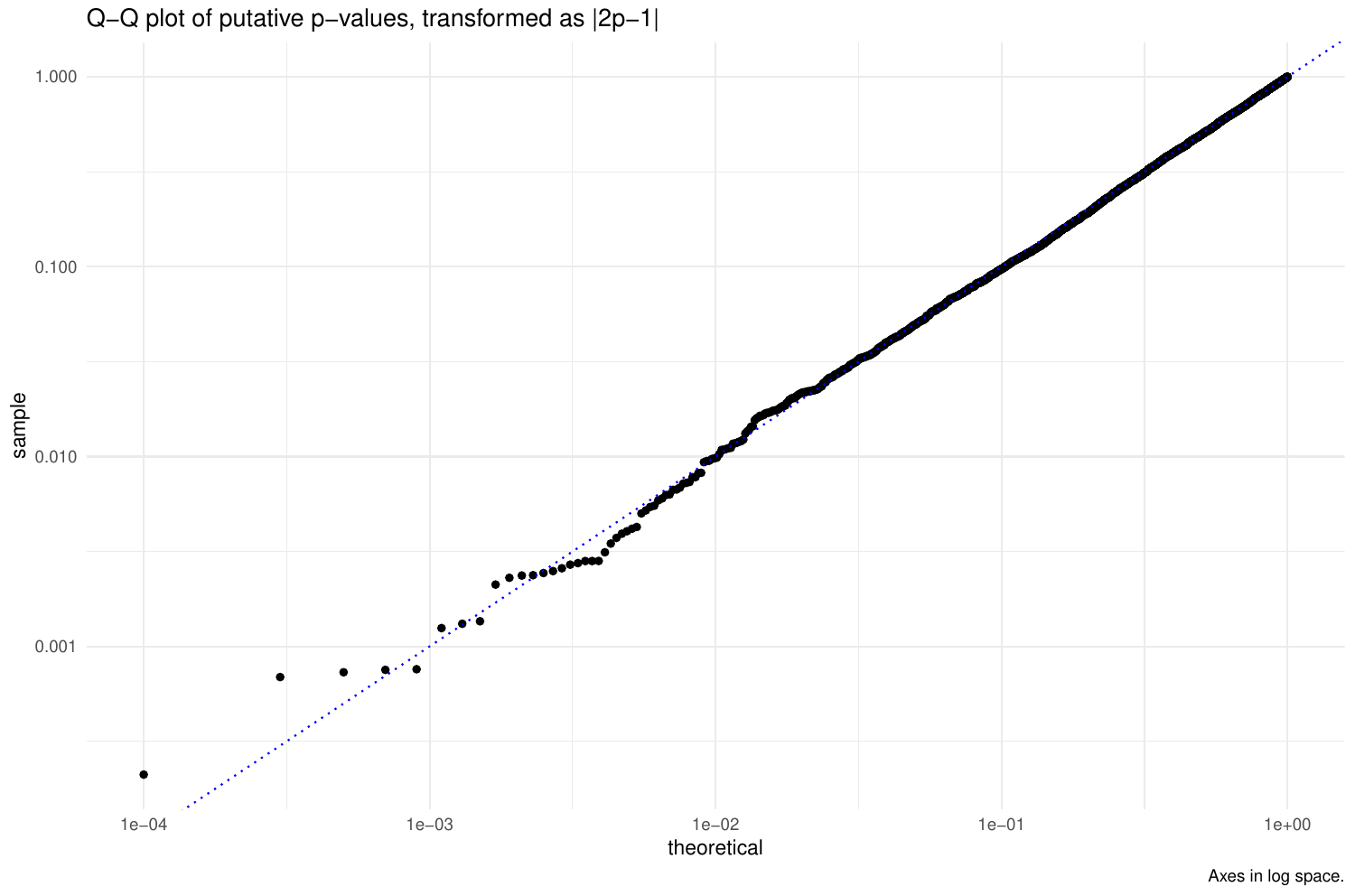} \caption{The computed p-values from 5,000 simulations are plotted against a uniform law, visually confirming that the p-values are nearly uniform. Simulations use the exact $\rho$ to compute the p-values via \code{ptukey}. We transform the p-values and plot $\abs{2p - 1}$ in log-log space to emphasize the tails. The points show little deviation from the plotted $y=x$ line. }\label{fig:null_basic_pp_plot}
\end{figure}

\end{knitrout}


\clearpage

\paragraph{Varying \ssiz and \nstrat:}

Next we perform the same kind of simulations, but vary the number of days observed in each
simulation, \ssiz, as well as the number of different assets, \nstrat.
We let the former vary 
from 20 to 1,280
measured in days, 
and the latter vary 
from 8 to 32.
We take
$\RMAT=\makerho{\rho}{\wrapParens{1-\rho}},$ 
for $\rho=0.8$ and set the \txtSNR to $1\yrto{-\halff}$.
We assume $252$ days per year for annualizing the \txtSR.
For each set of $50,000$ simulations,
we compute the empirical type I rate at the nominal $0.05$ level 
by comparing the range to the HSD cutoff.
We tabulate rejections using both the $df=\infty$ and $df=\ssiz-1$ cutoffs.

We plot that type I rate against $\ssiz$ in \figref{null_day_scan_plot}
for the different values of $\nstrat$. 
For the $df=\infty$ cutoff, the procedure is apparently
anticonservative, yielding too many type I errors, when the sample size is small
and the number of assets is large. 
For the $df=\ssiz-1$ cutoff, however, the nominal type I rate is approximately achieved.


\begin{knitrout}\small
\definecolor{shadecolor}{rgb}{0.969, 0.969, 0.969}\color{fgcolor}\begin{figure}[h]
\includegraphics[width=0.975\textwidth,height=0.646\textwidth]{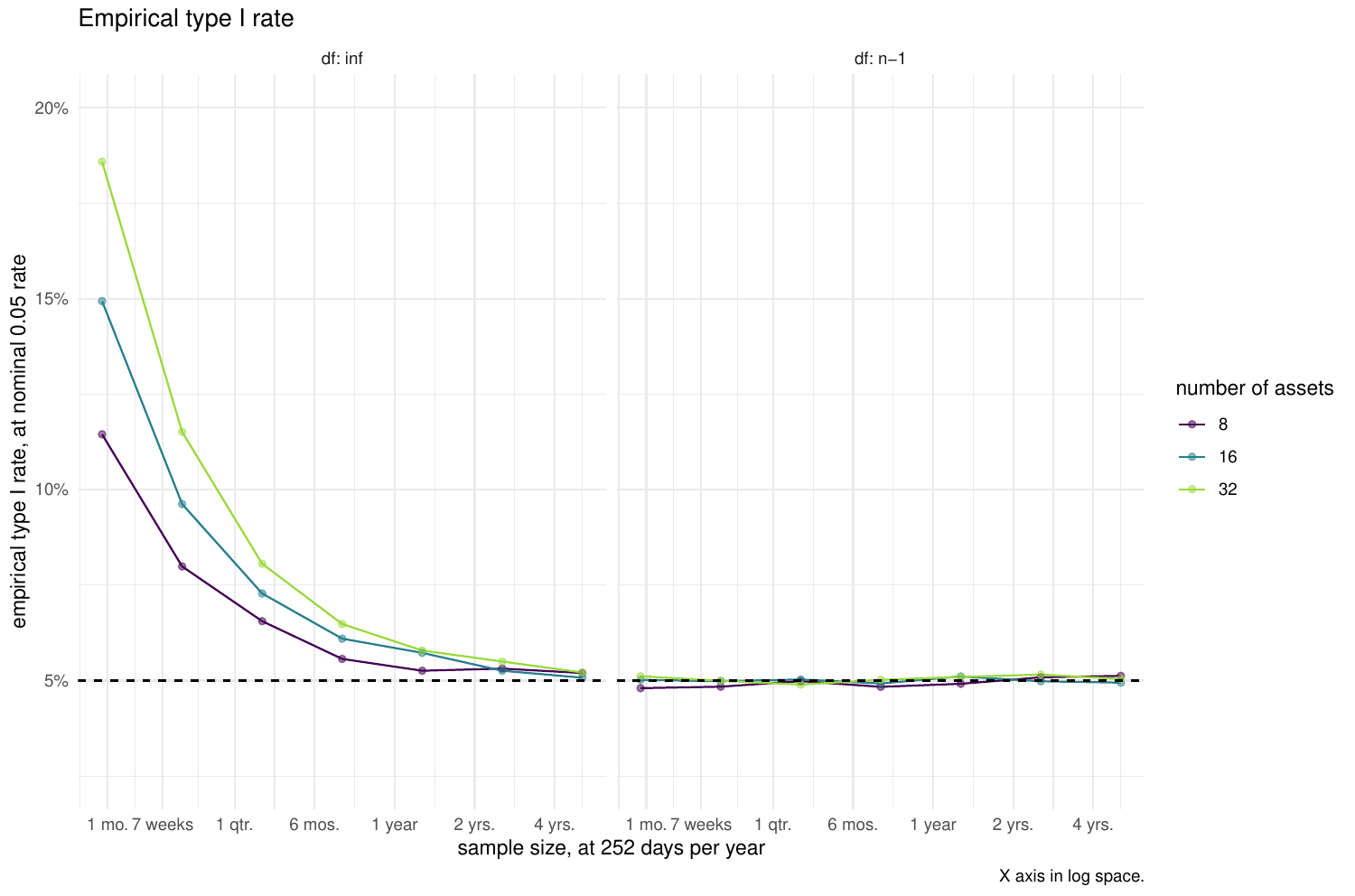} \caption[The empirical type I rate at the nominal 0.05 level is plotted against the number of days in each simulation, for different values of $\nstrat$]{The empirical type I rate at the nominal 0.05 level is plotted against the number of days in each simulation, for different values of $\nstrat$. Simulations use the exact $\rho$ to perform the hypothesis test. The two facets show rejection rates under the $df=\ssiz-1$ and $df=\infty$ cutoffs. When using the $df=\infty$ cutoff, the test is anti-conservative for the ``large \nstrat, small \ssiz'' case, but the nominal rate is nearly achieved for the $df=\ssiz-1$ cutoff.}\label{fig:null_day_scan_plot}
\end{figure}

\end{knitrout}

\paragraph{Varying $\rho$:}

We next perform the same simulations under the null, 
with $\RMAT=\makerho{\rho}{\wrapParens{1-\rho}},$ 
but scanning through $\rho$.
We set $\ssiz=1,008$ days, $\nstrat=16$,
and set the \txtSNR to $1\yrto{-\halff}$.
We compute the empirical type I rate at the nominal $0.05$ level for
each set of $50,000$ simulations.
We plot that type I rate against $\rho$ in \figref{null_rho_scan_plot}.
For these values of $\ssiz, \nstrat$, the procedure achieves near nominal
type I rate, and does not vary in a systematic way with $\rho$.

\begin{knitrout}\small
\definecolor{shadecolor}{rgb}{0.969, 0.969, 0.969}\color{fgcolor}\begin{figure}[h]
\includegraphics[width=0.975\textwidth,height=0.646\textwidth]{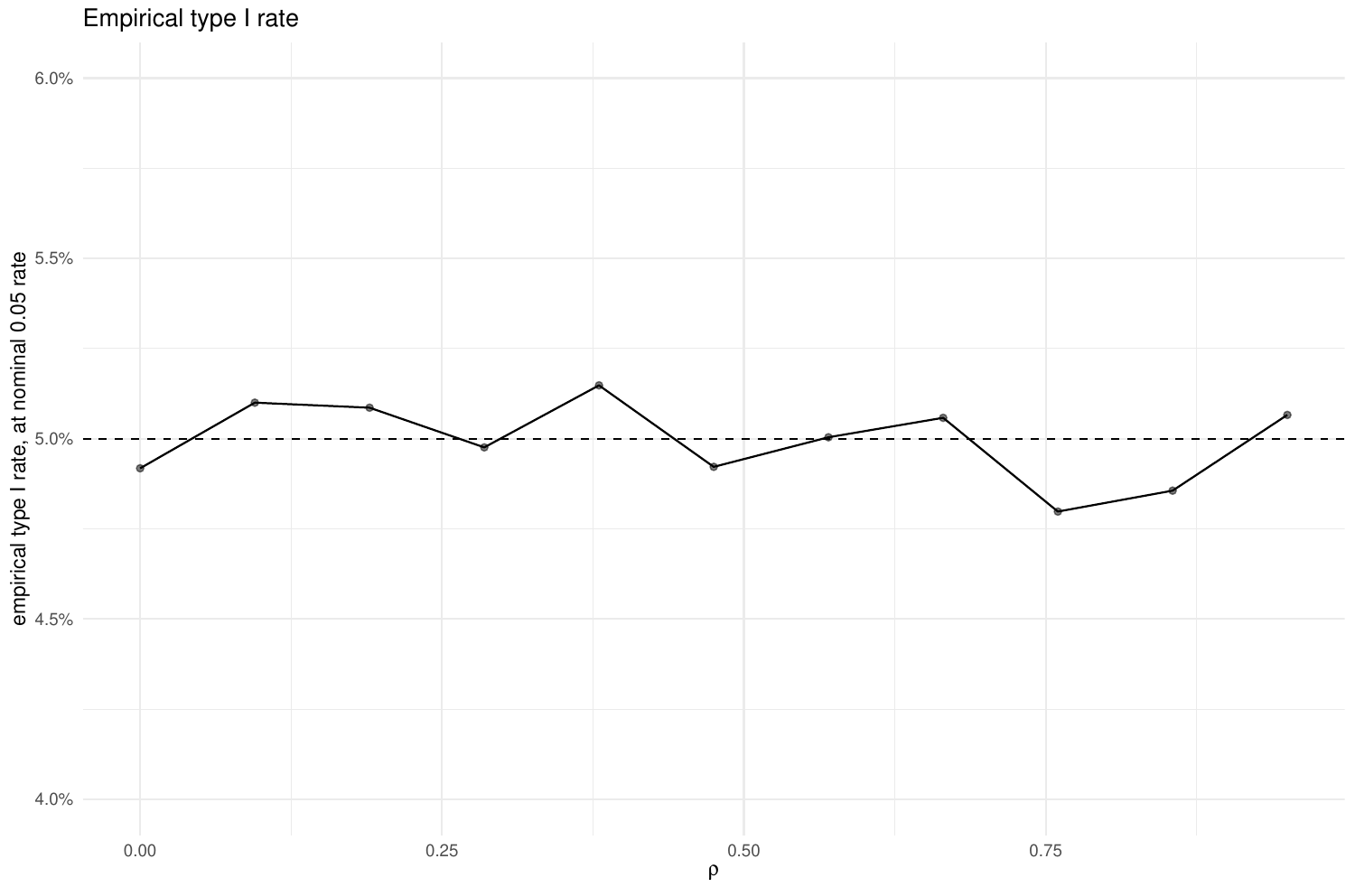} \caption[The empirical type I rate at the nominal 0.05 level is plotted against the correlation, $\rho$]{The empirical type I rate at the nominal 0.05 level is plotted against the correlation, $\rho$. Simulations use the exact $\rho$ to perform the hypothesis test, and the $df=\ssiz-1$ cutoff. }\label{fig:null_rho_scan_plot}
\end{figure}

\end{knitrout}

\paragraph{Feasible Estimator, Varying $\rho$:}

In the simulations above we have used the actual $\rho$ in computing
the threshold for rejection of the null.
We repeat the experiments using a \emph{feasible} test where we
esimate $\rho$ from the sample. 
We compute the correlation of returns, then take the median value of 
the upper triangle of the correlation matrix.

In the first set of simulations, the true correlation matrix follows
$\RMAT=\makerho{\rho}{\wrapParens{1-\rho}}.$
We set $\ssiz=1,008$ days, $\nstrat=16$,
and set the \txtSNR to $1\yrto{-\halff}$.
We compute the empirical type I rate at the nominal $0.05$ level for
each set of $50,000$ simulations.
We plot that type I rate against $\rho$ in \figref{null_feas_rho_scan_plot}.
For these values of $\ssiz, \nstrat$, the procedure achieves near nominal
type I rate, and does not appear to suffer from having estimated the $\rho$.
In fact the plot greatly resembles \figref{null_rho_scan_plot} where
we have used the actual $\rho$.

\begin{knitrout}\small
\definecolor{shadecolor}{rgb}{0.969, 0.969, 0.969}\color{fgcolor}\begin{figure}[h]
\includegraphics[width=0.975\textwidth,height=0.646\textwidth]{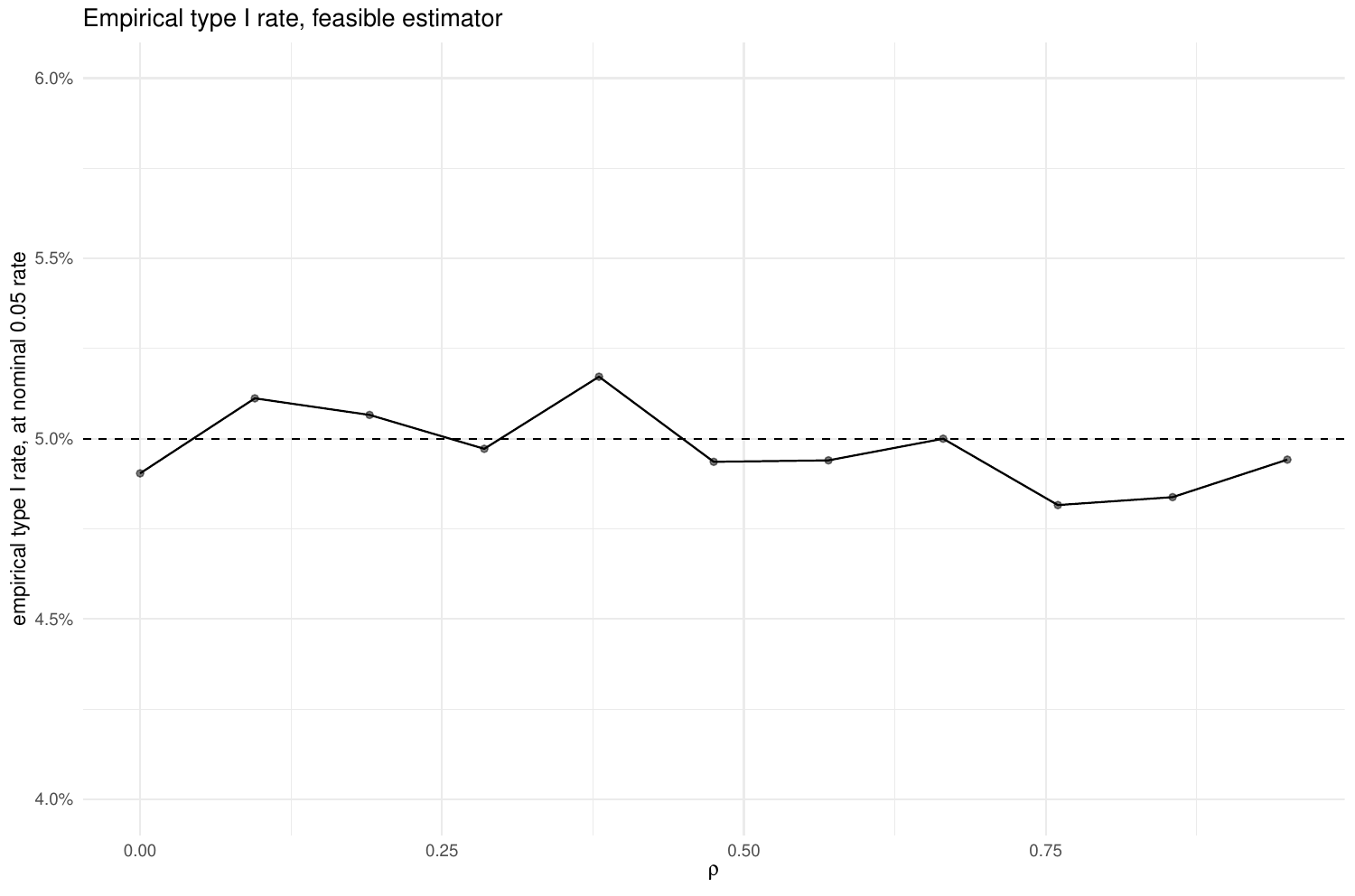} \caption[The empirical type I rate at the nominal 0.05 level is plotted against the correlation, $\rho$]{The empirical type I rate at the nominal 0.05 level is plotted against the correlation, $\rho$. Simulations use an \emph{estimated} $\rho$ to perform the hypothesis test, and the $df=\ssiz-1$ cutoff. }\label{fig:null_feas_rho_scan_plot}
\end{figure}

\end{knitrout}


\paragraph{Feasible Estimator, Misspecified Model, Varying $\rho$:}

We repeat those simulations, estimating the $\rho$ from the sample,
but now we let the correlation matrix take an ``AR(1)'' structure.
That is, we let $\RHAT[i,j] = \rho^{\abs{i-j}}$, and vary $\rho$.
Again we have 
$\ssiz=1,008$ days, 
$\nstrat=16$,
the \txtSNR is equal to $1\yrto{-\halff}$.
We compute the empirical type I rate at the nominal $0.05$ level for
each set of $50,000$ simulations.

For these simulations, we also record the type I rate when the $\rho$ is
not estimated, but instead assumed to be $0$.
Given that $\rho=0$ forms a kind of `stochastic lower bound', we expect
that the procedure will be anti-conservative when performed this way.
Indeed we see in the plot that the empirical type I rate
decreases to zero in increasing $\rho$.
For the case where we take the median sample correlation as the estimate
of $\rho$, the procedure is somewhat conservative for small $\rho$,
then anti-conservative for large $\rho$. This is not surprising:
for large $\rho$, the median element of \RHAT will be fairly
large, but the correlation among assets is somewhat weak. 
A more robust heuristic for estimating the $\rho$ is needed.

\begin{knitrout}\small
\definecolor{shadecolor}{rgb}{0.969, 0.969, 0.969}\color{fgcolor}\begin{figure}[h]
\includegraphics[width=0.975\textwidth,height=0.646\textwidth]{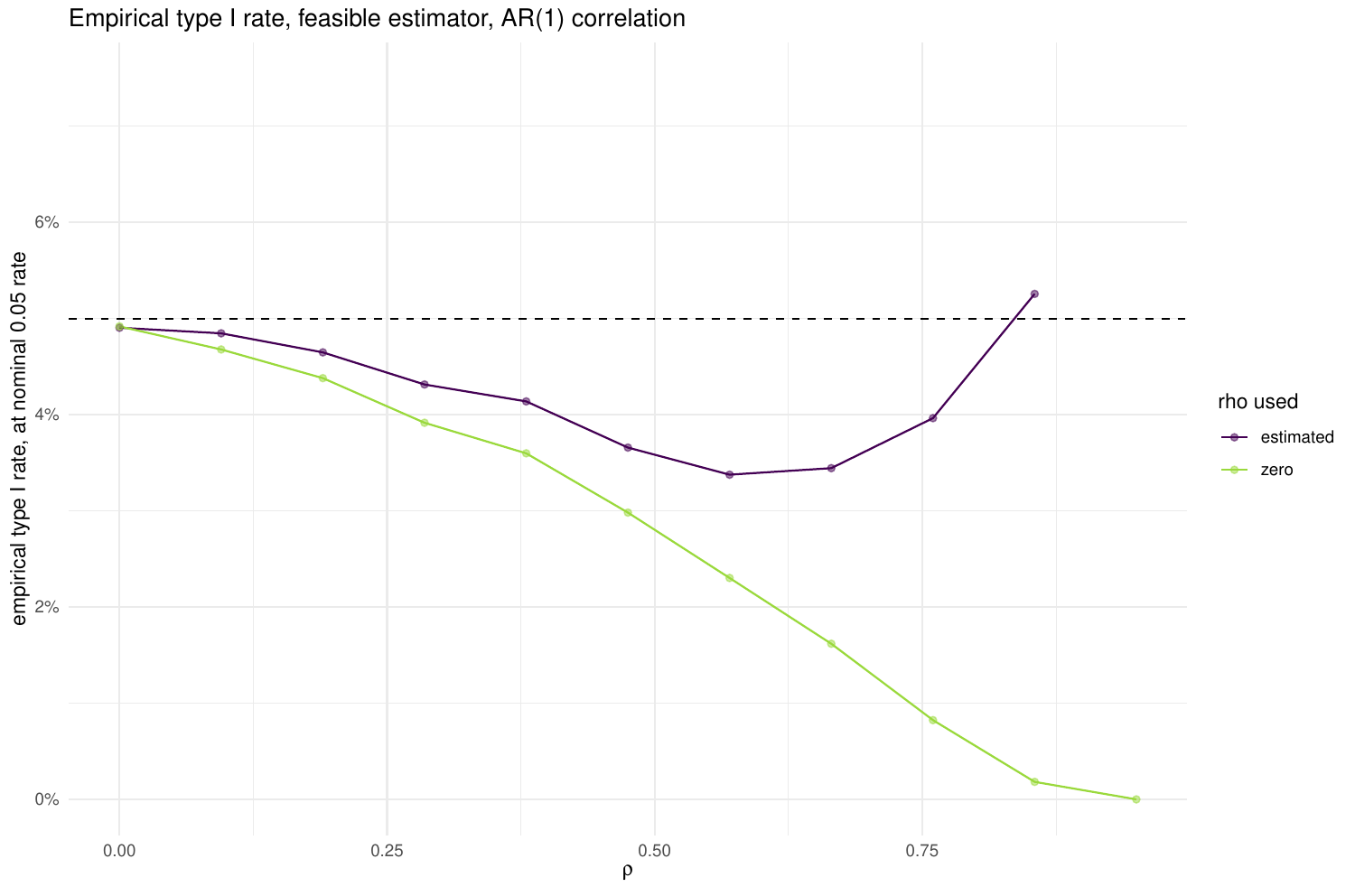} \caption[The empirical type I rate at the nominal 0.05 level is plotted against the correlation, $\rho$, for simulations where the correlation follows an AR(1) structure]{The empirical type I rate at the nominal 0.05 level is plotted against the correlation, $\rho$, for simulations where the correlation follows an AR(1) structure. Simulations use an \emph{estimated} $\rho$ to perform the hypothesis test, and the $df=\ssiz-1$ cutoff. We include separate lines for the cases where $\rho$ is estimated, and for where it is assumed to equal $0$. For the estimated $\rho$, the procedure is conservative for small and moderate $\rho$, but anticonservative for $\rho$ near 1. When $\rho=0$ is assumed, the procedure is increasingly conservative in $\rho$. }\label{fig:null_feas_ar1_rho_scan_plot}
\end{figure}

\end{knitrout}



\clearpage

\subsection{Simulations under the alternative}

We next perform the same simulations under the alternative.
It is somewhat difficult to quantify the power of this procedure
because the procedure can reject multiple nulls for a given
experiment. 
Indeed, in the simulations under the null above, we analyzed
the rate of \emph{any} rejections for the multiple comparisons performed
in a single simulation.

\paragraph{Under the alternative, one good: }

In the first set of simulations we let \pvsnr have a single non-zero
value, call it \psr, and vary that \psr.
We then compute, as the `range', the \txtSR of the single good asset
minus the minimum \txtSR of the $\nstrat - 1$ remaining assets.
Because we are only testing $\nstrat - 1$ comparisons, rather than
${\nstrat \choose 2}$, we expect to see fewer than the nominal type I 
rate when $\psr=0$. Moreover, we are performing a one-sided test.
As such it may be more natural to compare the rejection rate to
$\typeI/2$.

We take $\RMAT=\makerho{\rho}{\wrapParens{1-\rho}},$ letting
$\rho$ vary from 0 to 0.9;
we set $\ssiz=1,008$ days, 
$\nstrat=16$,
and let \psnr vary 
from $0\yrto{-\halff}$ 
to $1.5\yrto{-\halff}$.

We compute the rejection rate at the nominal $0.05$ level for
each set of $10,000$ simulations.
We plot that (true) rejection rate against $\psnr$ in \figref{alt_sim_onehi_scan_plot}.
For these values of $\ssiz, \nstrat$, the procedure is fairly weak,
only achieving power of one half for large \psnr or highly correlated
assets.
It is not surprising that the power is increasing in $\rho$: 
one expects less spread among the assets for higher $\rho$,
thus a true difference in \txtSNR is more easily detected.
This same effect is visible in the paired test for
equality of \txtSNRs.  \cite{pav_ssc}

\begin{knitrout}\small
\definecolor{shadecolor}{rgb}{0.969, 0.969, 0.969}\color{fgcolor}\begin{figure}[h]
\includegraphics[width=0.975\textwidth,height=0.646\textwidth]{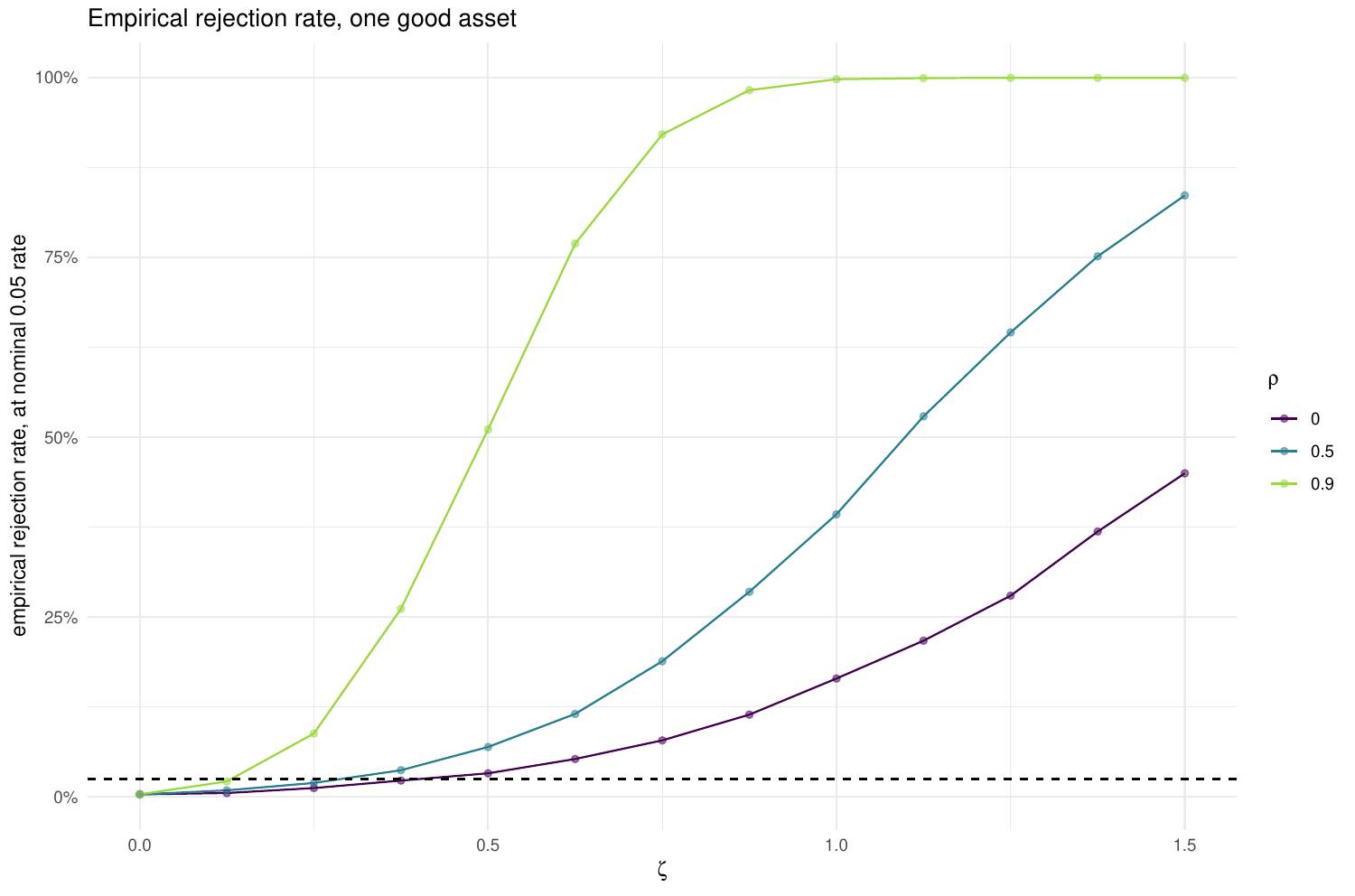} \caption[The empirical rejection rate at the nominal 0.05 level is plotted against \psnr]{The empirical rejection rate at the nominal 0.05 level is plotted against \psnr. The population consists of one good asset with \txtSNR equal to \psnr, and the remainder with zero \txtSNR. Rejection is based on the \txtSR of the single good asset minus the minimum \txtSR of the remaining assets. Simulations use the exact $\rho$ to perform the hypothesis test, and the $df=\ssiz-1$ cutoff. We plot a horizontal line at half the nominal type I rate, 0.025, because we are performing a one-sided test. }\label{fig:alt_sim_onehi_scan_plot}
\end{figure}

\end{knitrout}

\paragraph{Under the alternative, half good: }

We repeat those experiments, but set half the $16$ assets
to have \txtSNR equal to \psnr, and the rest to have zero \txtSNR.
We compute, as the `range', the maximum \txtSR of the good assets
minus the minimum \txtSR of the $\nstrat - 1$ remaining assets.
We are effectively testing $\wrapParens{\nstrat / 2}^2$ comparisons,
rather than
${\nstrat \choose 2}$, so we expect to see fewer than the nominal type I 
rate when $\psr=0$. 

As above 
we take $\RMAT=\makerho{\rho}{\wrapParens{1-\rho}},$ let
$\rho$ vary from 0 to 0.9,
$\ssiz=1,008$ days, 
$\nstrat=16$,
and let \psnr vary 
from $0\yrto{-\halff}$ 
to $1.5\yrto{-\halff}$.

We compute the rejection rate at the nominal $0.05$ level for
each set of $10,000$ simulations.
We plot that (true) rejection rate against $\psnr$ in \figref{alt_sim_halhi_scan_plot}.
For these values of $\ssiz, \nstrat$, the procedure is again
fairly underpowered, with higher power for more correlated
assets.

\begin{knitrout}\small
\definecolor{shadecolor}{rgb}{0.969, 0.969, 0.969}\color{fgcolor}\begin{figure}[h]
\includegraphics[width=0.975\textwidth,height=0.646\textwidth]{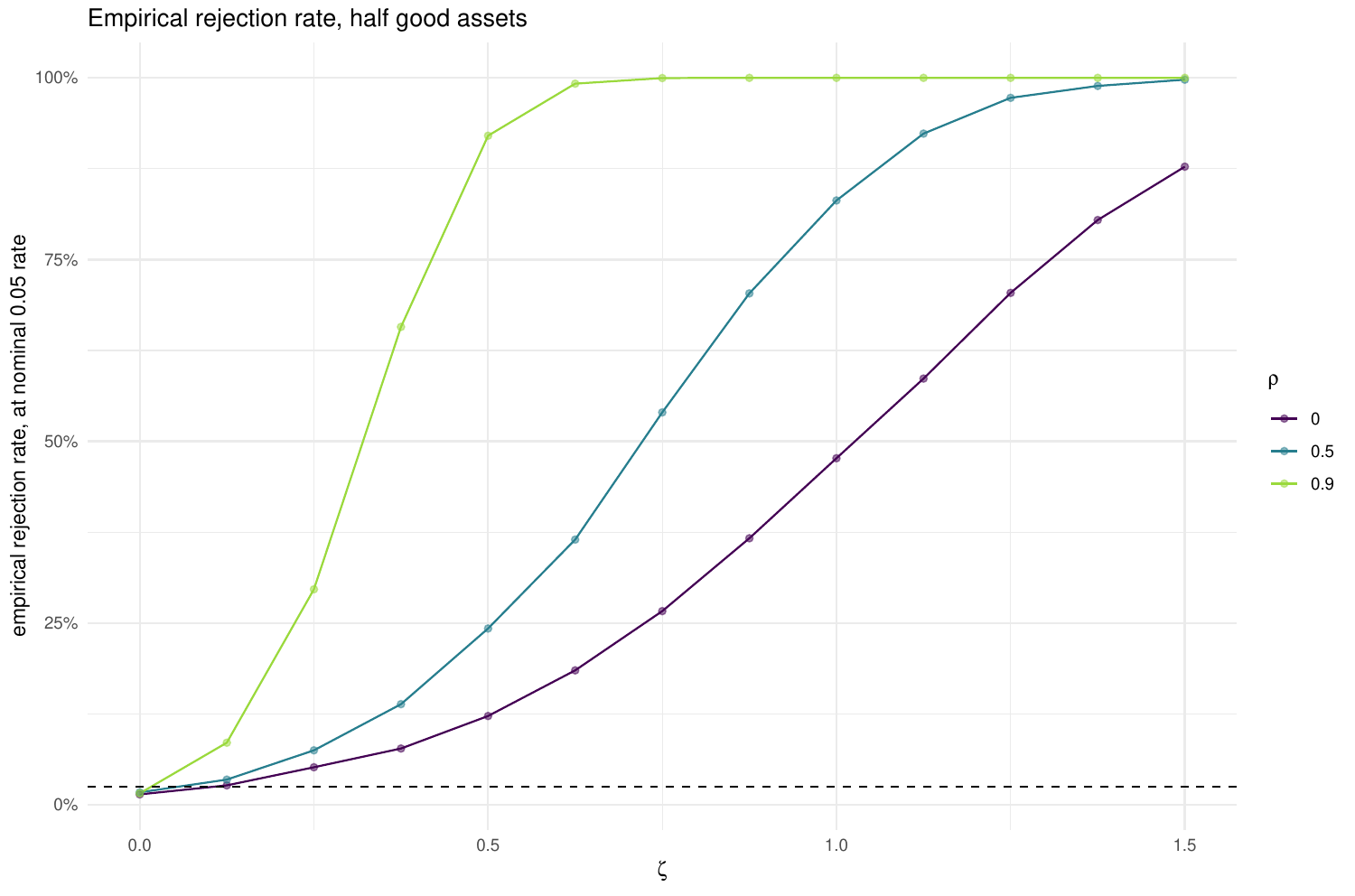} \caption[The empirical rejection rate at the nominal 0.05 level is plotted against \psnr]{The empirical rejection rate at the nominal 0.05 level is plotted against \psnr. The population consists of half good assets with \txtSNR equal to \psnr, and the remainder with zero \txtSNR. Rejection is based on the maximum \txtSR of the good assets minus the minimum \txtSR of the remaining assets. Simulations use the exact $\rho$ to perform the hypothesis test, and the $df=\ssiz-1$ cutoff. We plot a horizontal line at half the nominal type I rate, 0.025, because we are performing a one-sided test. }\label{fig:alt_sim_halhi_scan_plot}
\end{figure}

\end{knitrout}

\clearpage

\subsection{Real Assets}

We now apply the technique to real asset returns.

\paragraph{Five Industry Portfolios:}

We consider the 5 industry portfolios, whose returns are computed and distributed by French.  \cite{ind_5_def} 
The dataset consists of 1080 months of returns,
from Jan 1927 to Dec 2016.
The returns are highly correlated, and the correlation matrix is likely
well modeled by the form 
$\makerho{\rho}{\wrapParens{1-\rho}},$
with $\rho$ estimated as approximately $0.8$.
The \txtSRs range from $0.487\yrto{-\halff}$ for Other to
$0.664\yrto{-\halff}$ for Healthcare.

First we perform the hypothesis test of equality of \txtSNRs, as proposed by 
Wright \etal \cite{wright2014}
We compute a statistic of $12$ which should be distributed as a
$\chi^2\wrapParens{4}$ under the null. \cite{pav_ssc}
This corresponds to a p-value of $0.017$, and we reject the null
of equality of all \txtSNRs.

Using the $df=\ssiz-1$ formulation and the estimated $\rho$, we compute
$HSD = 0.181\yrto{-\halff}$ for $\typeI = 0.05$, 
and narrowly reject the equality of \txtSNRs for Other and Healthcare.
In \figref{mind5_lolly_plot}, we plot these \txtSRs, along with error bars
at plus and minus one $HSD$.

\begin{knitrout}\small
\definecolor{shadecolor}{rgb}{0.969, 0.969, 0.969}\color{fgcolor}\begin{figure}[h]
\includegraphics[width=0.975\textwidth,height=0.390\textwidth]{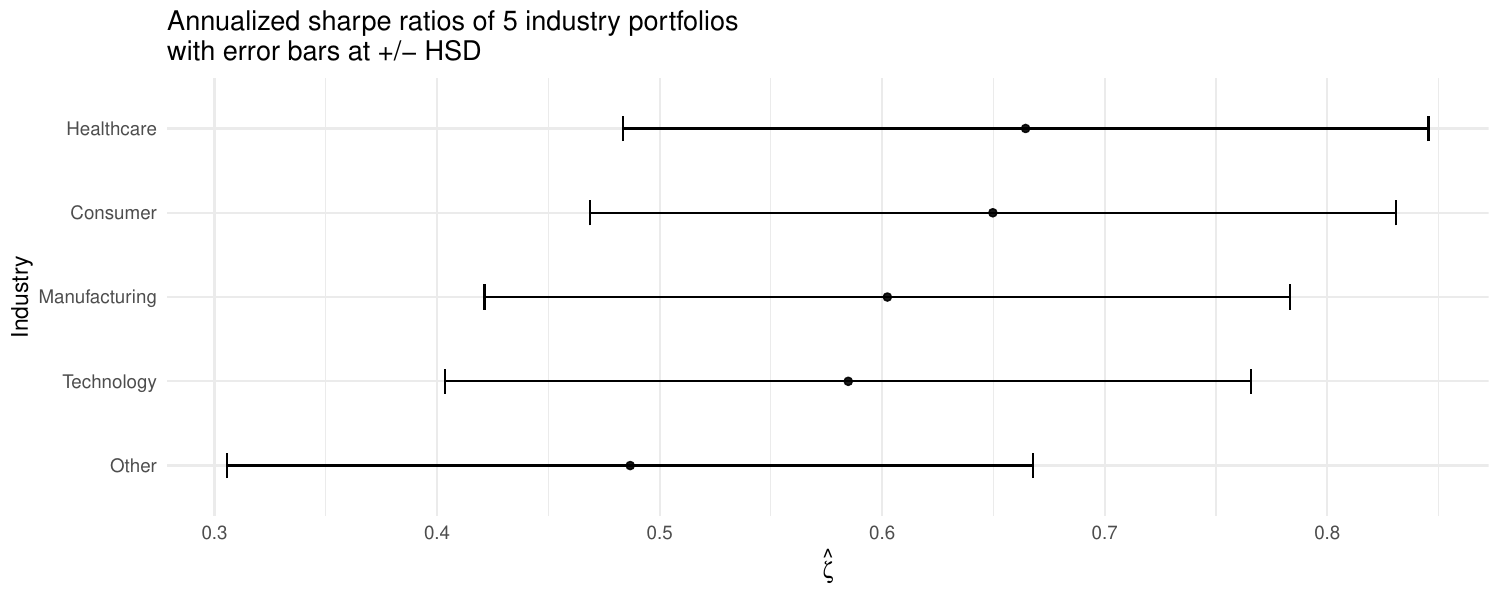} \caption[The annualized \txtSR of French's 5 industry portfolios are plotted, as computed on monthly returns from Jan 1927 to Dec 2016]{The annualized \txtSR of French's 5 industry portfolios are plotted, as computed on monthly returns from Jan 1927 to Dec 2016. We plot error bars at $\pm HSD$ for $\typeI=0.05$. We narrowly reject equality of the \txtSNR of Other and Healthcare. }\label{fig:mind5_lolly_plot}
\end{figure}

\end{knitrout}

\paragraph{Sharpe's 34 Mutual Funds: }

We consider the returns of the 34 mutual funds described by Sharpe in his
original paper.  \cite{Sharpe:1966} 
We transcribed the annualized percent return and standard deviation values from 
Sharpe's Table I.
In his paper, Sharpe computed the ``reward-to-variability ratio'' of each using a fixed
rate of $3\%$;
however, we compute the \txtSR without subtracting a fixed rate.
The \txtSRs range from $0.549\yrto{-\halff}$ for Incorporated Investors to
$1.087\yrto{-\halff}$ for American Business Shares.

We do not have access to the series of returns, and cannot estimate the correlation
structure. 
Somewhat optimistically we make the wild guess $\rho=0.85$.
Based on this value, and setting $\typeI=0.05$, we
compute $HSD = 0.68\yrto{-\halff}$,
and we fail to reject the null hypothesis that all \txtSNRs are equal.
In \figref{sr34_lolly_plot}, we plot these \txtSRs, along with error bars
at $\pm HSD$.
Given the lack of separation of the funds, it is curious that
Sharpe found correlation between the in-sample and out-of-sample
\txtSRs of his funds. \cite{Sharpe:1966}

\begin{knitrout}\small
\definecolor{shadecolor}{rgb}{0.969, 0.969, 0.969}\color{fgcolor}\begin{figure}[h]
\includegraphics[width=0.975\textwidth,height=0.812\textwidth]{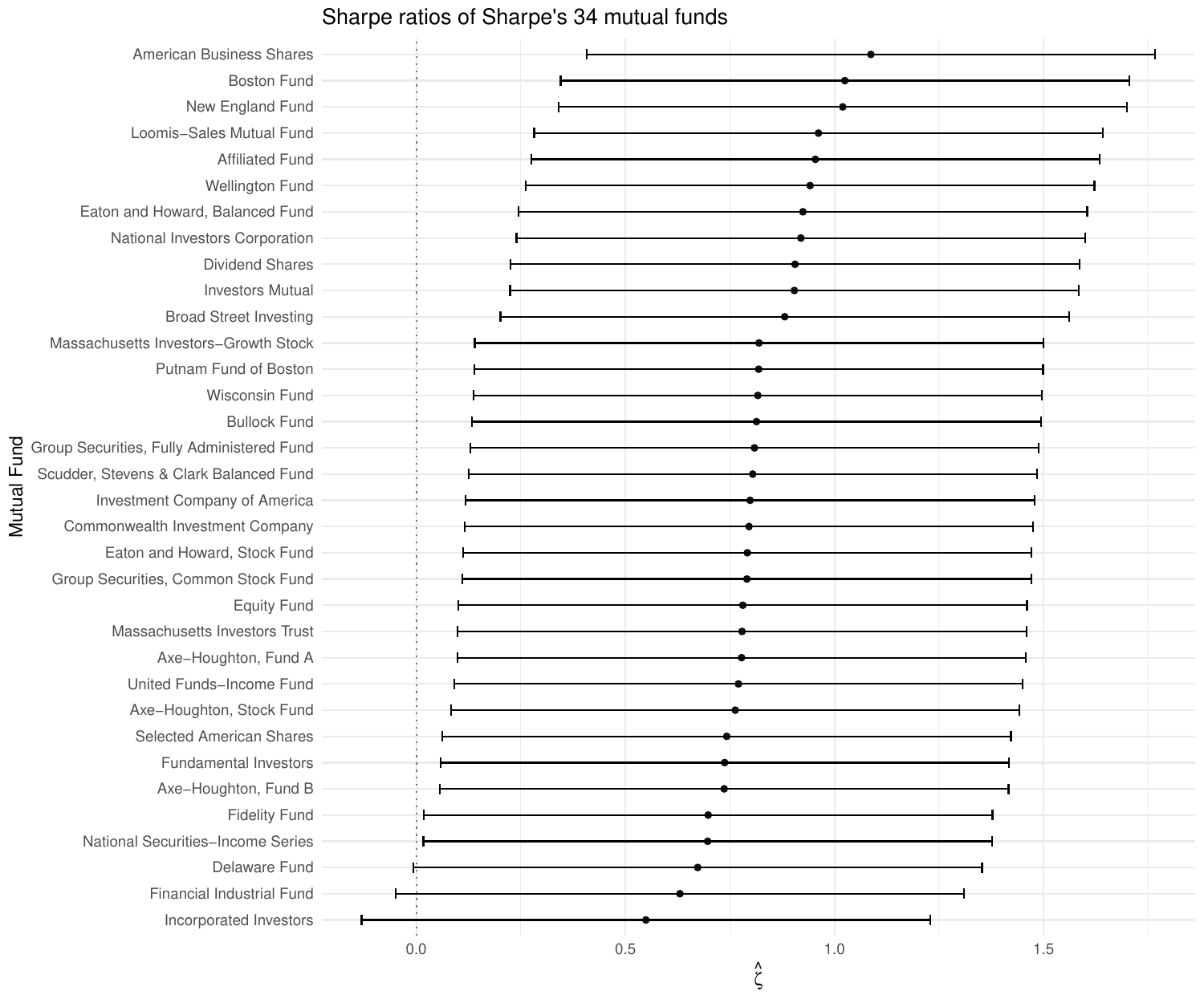} \caption{The annualized \txtSR of Sharpe's 34 mutual funds are plotted. Returns are from the decade 1954-1963. \cite{Sharpe:1966} We plot error bars at $\pm HSD$. We fail to reject the nulls that all pairwise differences are equal. }\label{fig:sr34_lolly_plot}
\end{figure}

\end{knitrout}

\section{Future Work}

A number of issues remain outstanding:

\begin{enumerate}
\item The heuristic use of the $df=\ssiz-1$ cutoff requires theoretical
  justification.
\item A stochastic inequality like Slepian's lemma for ranges
  would allow one to apply the test using a lower bound $\rho$ to
  achieve maximum type I rate.
\item Should we expect the Tukey HSD cutoff and the Bonferroni Cutoff to 
  be nearly equal, or will one dominate the other under
  certain conditions?
\item Can we quantify the power of the test? 
\item 
  Though we suspect it cannot, 
  can the power of the test be improved? 
\end{enumerate}


\bibliographystyle{plainnat}
\bibliography{common}

@article{Johnson:1940,
	author = {Johnson, N. L. and Welch, B. L.},
	day = {1},
	doi = {10.1093/biomet/31.3-4.362},
	journal = {Biometrika},
	month = "mar",
	number = {3-4},
	pages = {362--389},
	title = {Applications of the non-central t-distribution},
	url = {http://dx.doi.org/10.1093/biomet/31.3-4.362},
	volume = {31},
	year = {1940}
}

@Article{Sharpe:1966,
	author={Sharpe, William F.},
	title={Mutual Fund Performance},
	journal={Journal of Business},
	year=1965,
	volume={39},
	number={},
	pages={119},
	month={},
	url={http://ideas.repec.org/a/ucp/jnlbus/v39y1965p119.html}
}

@article{lo2002,
	type={Accepted Paper Series},
	title={The Statistics of {S}harpe Ratios},
	author={Lo, Andrew W.},
	location={http://ssrn.com/paper=377260},
	journal = {Financial Analysts Journal},
	volume = 58,
	number = 4,
	month = {July/August},
	year = "2002",
	url={http://ssrn.com/paper=377260},
	language={English}
}

@article{jobsonkorkie1981,
     jstor_articletype = {research-article},
     title = {Performance Hypothesis Testing with the {S}harpe and {T}reynor Measures},
     author = {Jobson, J. D. and Korkie, Bob M.},
     journal = {The Journal of Finance},
     jstor_issuetitle = {},
     volume = {36},
     number = {4},
     jstor_formatteddate = {Sep., 1981},
     pages = {pp. 889-908},
     url = {http://www.jstor.org/stable/2327554},
     ISSN = {00221082},
     language = {English},
     year = {1981},
     publisher = {Wiley for the American Finance Association},
     copyright = {Copyright © 1981 American Finance Association},
}

@Manual{Rlang,
	title = {R: A Language and Environment for Statistical Computing},
	author = {{R Core Team}},
	organization = {R Foundation for Statistical Computing},
	address = {Vienna, Austria},
	year = {2015},
	note         = {{ISBN} 3-900051-07-0},
	url = {http://www.R-project.org/},
}

@article{CambridgeJournals:4493808,
	author = {Miller, Robert E. and Gehr, Adam K.},
	title = {Sample Size Bias and {S}harpe's Performance Measure: A Note},
	journal = {Journal of Financial and Quantitative Analysis},
	volume = {13},
	number = {05},
	pages = {943--946},
	year = {1978},
	doi = {10.2307/2330636},
	badURL = {http://dx.doi.org/10.1017/S0022109000014216},
	URL = {https://doi.org/10.2307/2330636},
	eprint = {http://journals.cambridge.org/article_S0022109000014216}
}

@article{Leung2008,
	title = {On testing the equality of multiple {S}harpe ratios, with application on the evaluation of {iShares}},
	author = {Leung, Pui-Lam and Wong, Wing-Keung},
	journal = {Journal of Risk},
	volume = 10,
	number = 3,
	year = 2008,
	pages = {15--30},
	url = {http://www.risk.net/digital_assets/4760/v10n3a2.pdf},
	url2 = {http://ied.hkbu.edu.hk/publications/fdp/FDP200803.pdf},
	url3 = {http://papers.ssrn.com/sol3/papers.cfm?abstract_id=907270}
}

@article{wright2014,
  author = {Wright, John Alexander and Yam, Sheung Chi Phillip and Yung, Siu Pang},
  description = {Risk.net - A test for the equality of multiple Sharpe ratios},
  journal = {Journal of Risk},
  number = 4,
  title = {A test for the equality of multiple {S}harpe ratios},
  url = {http://www.risk.net/journal-of-risk/journal/2340044/latest-issue-of-the-journal-of-risk-volume-16-number-4-2014},
  volume = 16,
  year = 2014
}

@book{bain1992introduction,
  title={Introduction to Probability and Mathematical Statistics},
  author={Bain, L. J. and Engelhardt, M.},
  isbn={9780534380205},
  lccn={91025923},
  series={Classic Series},
  url={http://books.google.com/books?id=MkFRIAAACAAJ},
  year={1992},
  publisher={Cengage Learning}
}

@misc{pav2019maxsharpe,
  author={Pav, Steven E.},
  title={Conditional inference on the asset with maximum {S}harpe ratio},
	url = {http://arxiv.org/abs/1906.00573},
	howpublished = {Privately Published},
  year={2019}
}

@book{pav_the_book,
	title={The {S}harpe Ratio: Statistics and Applications},
	Author={Pav, Steven E.},
	publisher={CRC Press},
  isbn={9781000442762},
	url={https://www.google.com/books/edition/The_Sharpe_Ratio/Zww8EAAAQBAJ},
	year=2021
}

@article{roySafety1952,
	 jstor_articletype = {research-article},
	 title = {Safety First and the Holding of Assets},
	 author = {Roy, A. D.},
	 journal = {Econometrica},
	 jstor_issuetitle = {},
	 volume = {20},
	 number = {3},
	 jstor_formatteddate = {Jul., 1952},
	 pages = {pp. 431-449},
	 url = {http://www.jstor.org/stable/1907413},
	 ISSN = {00129682},
	 language = {English},
	 year = {1952},
	 publisher = {The Econometric Society},
	 copyright = {Copyright © 1952 The Econometric Society},
	}

@misc{ind_5_def,
	title = {5 Industry Portfolios},
	author = {French, Kenneth},
	url = {http://mba.tuck.dartmouth.edu/pages/faculty/ken.french/Data_Library/det_5_ind_port.html},
	year = 2019,
	howpublished = {Privately Published}
}

@misc{pav_ssc,
	title={A Short {S}harpe Course},
	Author={Pav, Steven E.},
	year=2017,
	howpublished = {Privately Published},
	doi={10.2139/ssrn.3036276},
	dumbURL={https://dx.doi.org/10.2139/ssrn.3036276},
	url={https://papers.ssrn.com/sol3/papers.cfm?abstract_id=3036276}
}

@article{slepian1962one,
  title={The one-sided barrier problem for {G}aussian noise},
  author={Slepian, David},
  journal={Bell System Technical Journal},
  volume={41},
  number={2},
  pages={463-501},
  year={1962},
	doi={10.1002/j.1538-7305.1962.tb02419.x},
	url={https://onlinelibrary.wiley.com/doi/abs/10.1002/j.1538-7305.1962.tb02419.x},
	eprint = {https://onlinelibrary.wiley.com/doi/pdf/10.1002/j.1538-7305.1962.tb02419.x},
  publisher={Wiley Online Library}
}

@misc{yin2019stochastic,
    title={Stochastic Orderings of Multivariate Elliptical Distributions},
    author={Chuancun Yin},
    year={2019},
		url={https://arxiv.org/abs/1910.07158},
    eprint={1910.07158},
    archivePrefix={arXiv},
    primaryClass={math.ST}
}

@misc{zeitouni2015gaussian,
  title={{GAUSSIAN FIELDS} Notes for Lectures},
  author={Zeitouni, Ofer},
  year={2017},
	version={1.04g},
	howpublished={unpublished notes},
	url={http://www.wisdom.weizmann.ac.il/~zeitouni/notesGauss.pdf}
}

@book{bretz2016multiple,
  title={Multiple comparisons using {R}},
  author={Bretz, Frank and Hothorn, Torsten and Westfall, Peter},
  year={2016},
  publisher={Chapman and Hall/CRC},
	SEPnote={multiple comparisons and compact letter displays},
	url={http://www.ievbras.ru/ecostat/Kiril/R/Biblio_N/R_Eng/Bretz2011.pdf}
}

@article{tukey_hsd,
 ISSN = {0006341X, 15410420},
 URL = {http://www.jstor.org/stable/3001913},
 author = {John W. Tukey},
 journal = {Biometrics},
 number = {2},
 pages = {99--114},
 publisher = {[Wiley, International Biometric Society]},
 title = {Comparing Individual Means in the Analysis of Variance},
 doi={10.2307/3001913},
 volume = {5},
 year = {1949}
}

@article{OdehEvans,
 ISSN = {00359254, 14679876},
 URL = {http://www.jstor.org/stable/2347061},
 author = {R. E. Odeh and J. O. Evans},
 journal = {Journal of the Royal Statistical Society. Series C (Applied Statistics)},
 number = {1},
 pages = {96--97},
 doi={10.2307/2347061},
 publisher = {[Wiley, Royal Statistical Society]},
 title = {Algorithm {AS} 70: The Percentage Points of the Normal Distribution},
 volume = {23},
 year = {1974}
}







\end{document}